\documentclass[twocolumn]{aastex62}
\usepackage{natbib}

\graphicspath{{./}{}}

\received{Dec 21, 2018}
\revised{Jan 26, 2019}
\accepted{in ApJL Jan 30, 2019}

\shorttitle{Evolution of the Size--Mass Relation}
\shortauthors{Mowla et al.}

\begin{document}

\title{A mass-dependent slope of the galaxy size--mass relation out to $z\sim 3$: further evidence for a direct relation between median galaxy size and median halo mass}

\author{Lamiya Mowla}
\email{lamiya.mowla@yale.edu}
\affil{Astronomy Department, Yale University, New Haven, CT 06511, USA}

\author{Arjen van der Wel}
\affil{Sterrenkundig Observatorium, Universiteit Gent, Krijgslaan 281 S9, B-9000 Gent, Belgium}
\affil{Max-Planck-Institut f{\"u}r Astronomie, K{\"o}nigstuhl 17, D-69117, Heidelberg, Germany}

\author{Pieter van Dokkum}
\affil{Astronomy Department, Yale University, New Haven, CT 06511, USA}

\author{Tim B. Miller}
\affil{Astronomy Department, Yale University, New Haven, CT 06511, USA}

\begin{abstract}

We reassess the galaxy size-mass relation out to $z\sim 3$ using a new definition of size and a  sample of $>29,000$ galaxies from the 3D-HST, CANDELS, and COSMOS-DASH surveys.
Instead of the half-light radius $r_{50}$ we use $r_{80}$, the radius containing 80\,\% of the stellar light. We find that the $r_{80}$ -- $M_*$ relation
has the form of a broken power law, with a clear change of slope at a pivot mass $M_{\rm p}$. 
Below the pivot mass the relation is shallow ($r_{80}\propto M_*^{0.15}$) and above it it is steep ($r_{80}\propto M_*^{0.6}$). The pivot mass increases with redshift, from $\log(M_{\rm p}/{\rm M}_\odot)\approx 10.2$ at $z=0.4$ to $\log(M_{\rm p}/{\rm M}_{\odot})\approx 10.9$ at $z=1.7-3$. We compare these 
$r_{80}-M_*$ relations to the $M_{\rm halo}-M_*$ relations derived from galaxy-galaxy lensing, clustering analyses, and abundance matching techniques. Remarkably, the pivot stellar masses of both relations are consistent with each other at all redshifts, and the slopes are very similar both above and below the pivot when assuming $M_{\rm halo} \propto r_{80}^3$. The implied scaling factor to relate galaxy size to halo size is $r_{80} / R_{\rm vir} = 0.047$, independent of stellar mass and redshift.
From redshift 0 to 1.5, the pivot mass also coincides with the mass where the fraction of star-forming galaxies is 50\,\%, suggesting that the pivot mass reflects a transition from dissipational to dissipationless galaxy growth.
Finally, our results imply that the scatter in the stellar-to-halo mass ratio is relatively small for massive halos ($\sim 0.2$~dex for $M_{\rm halo}>10^{12.5} {\rm M}_\odot$).

\end{abstract}

\section{Introduction}

\begin{figure*}[htbp]
\centering
\includegraphics[width=\textwidth]{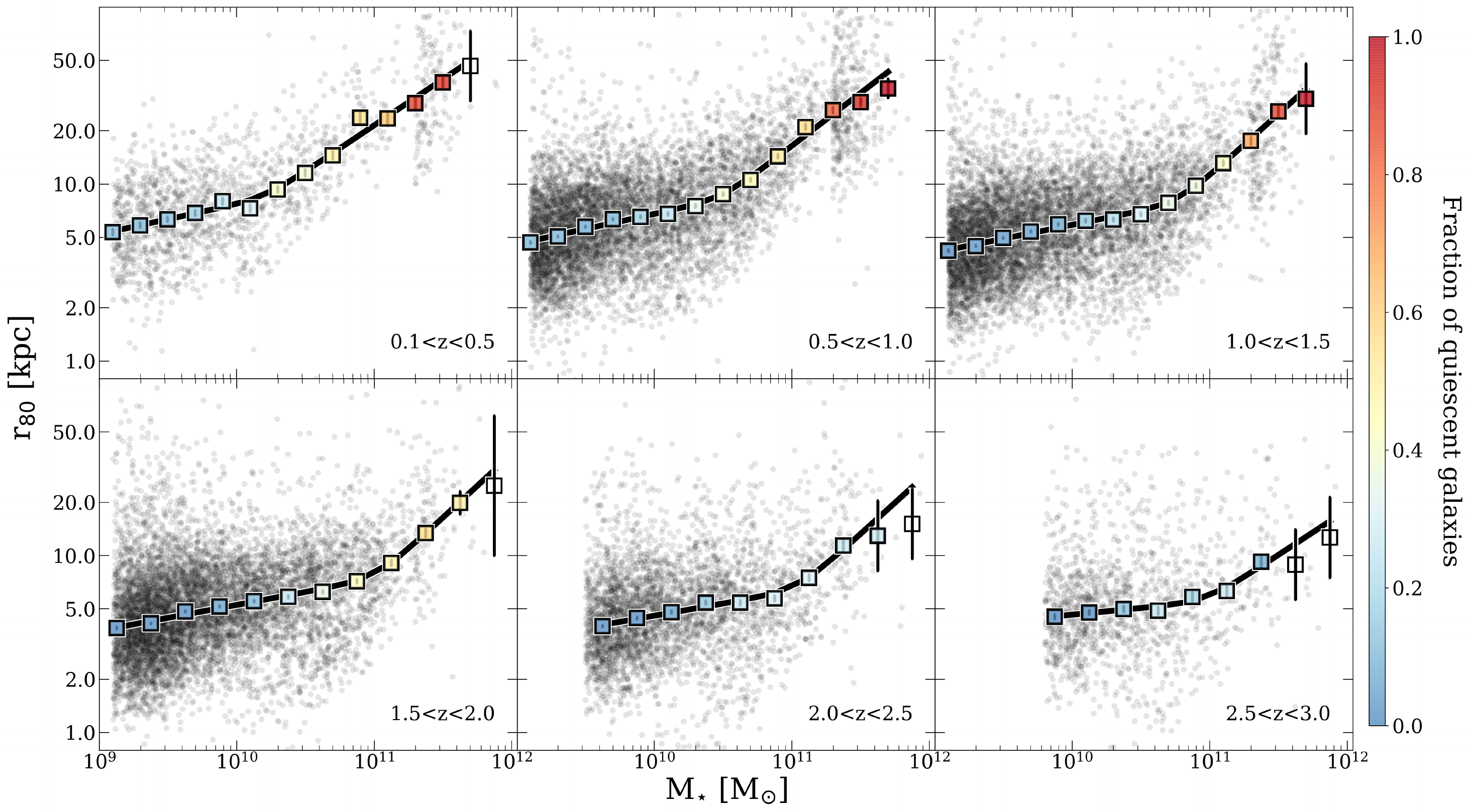}
\caption{Size-stellar mass distribution of galaxies at $0<z<3$ from \citet{VanderWel2014} and \citet{Mowla2018}. The squares show the median of $r_{\rm 80}$ in bins of $\log(M_\star/M_\odot)$, color-coded by the fraction of quiescent galaxies in the bin; rest-frame U−V and V−J color space was used to separate galaxies into star-forming and quiescent. Unfilled squares represent bins with less than 15 galaxies. Smoothly broken power law fits given by Eq. \ref{eq:power_law} to the median size--mass relation are shown by the black lines. } 
\label{fig:data_median_fit}
\end{figure*}

The size distribution of galaxies holds clues to their assembly history and the relationship with their dark matter halos \citep{Mo1997,Kravtsov2012,Jiang2018}. The sizes of galaxies are known to vary with stellar mass, star formation rate, and redshift and have been studied extensively  \citep[e.g.,][among many others]{Kormendy1977,Shen2003,Ferguson2003,Trujillo2006,Elmegreen2007,Williams2010,Ono2012,Bernardi2012,Mosleh2012,Carollo2013,VanderWel2014,Navarro2017,Kravtsov2018,Mowla2018}. 

One of the key result of these studies is that the two main classes of galaxies, star-forming and quiescent galaxies, follow very different size--mass relations. Hence it has been common practice to describe the size--mass distribution of galaxies separately for the two classes. It is usually defined by single power-law relation for each sub-population, with quiescent galaxies having a steeper relation than star forming ones \citep[e.g.,][]{Shen2003,VanderWel2014}. 
The interpretation of these results is a topic of debate; one possibility
is that star forming galaxies build up their stellar populations at all radii whereas quiescent galaxies mostly grow inside-out through accretion \citep[e.g.,][]{VanDokkum2015}.

In this \textit{Letter} we revisit the form of the size--mass relation out to $z=3$, using a large sample and an alternative size definition. This study is motivated by the availability of a new, large sample of distant galaxies with HST-measured sizes out to $z=3$ \citep{Mowla2018}, which extends to higher masses than previous studies \citep{VanderWel2014}. 

We find that the size--mass distribution of all galaxies is not well fit by a single power law but requires a change in slope at a pivot mass. We compare this to stellar-to-halo mass (SMHM) relations from the literature, assuming a conversion from size to virial radius \citep[e.g.,][]{Leauthaud2011,Moster2013,Behroozi2018}.
Our study extends earlier work at low redshift which found a steepening of the size--mass distribution at the high mass end for disk-dominated galaxies \citep[e.g.,][]{Shen2003,Dutton2010}, and theoretical work which suggested a constant scaling between the half-light radius and the virial radius of galaxies \citep{Kravtsov2012,Somerville2017, Huang2017,Jiang2018,Genel2018}.
We assume a flat $\Lambda$CDM cosmology with parameters $\Omega_m= 0.308$,$\Omega_b= 0.049$, h$ = H_0$/(100 km s$^{−1}$ Mpc$^{−1}$) $= 0.677$, $\sigma_8 = 0.823$ and $n_s = 0.96$ compatible with \textit{Planck} constraints \citep{PlanckCollaboration2015}.

\section{Data}

\subsection{Galaxy Sample}

The dataset we use is described in \citet{Mowla2018} and consists of the combination of two distinct samples. The first is from the CANDELS/3D-HST surveys. Sizes of over 28,000 galaxies with $M_{\star}>10^9 {\rm M}_{\odot}$ at $0<z<3$ were measured by \citet{VanderWel2014} from the 0.22 deg$^2$ CANDELS \citep{Koekemoer2011} imaging in the $H_{160}$, $J_{125}$ and $I_{814}$ bands. Spectroscopic and photometric redshifts, stellar masses, and rest-frame properties were measured by \citet{Skelton2014} using the extensive 3D-HST multi-wavelength data.

The area of the CANDELS/3D-HST fields is insufficient to properly sample the massive end of the luminosity function. This situation has been mitigated by the completion of COSMOS-DASH survey, which tripled the area surveyed by HST in the near-IR. COSMOS-DASH has enabled us to extend the size--mass study to higher masses at $1.5<z<3.0$ \citep{Mowla2018}. Sizes of 162 galaxies with $M_{\star}>2\times10^{11}{\rm M}_{\odot}$ at z$>$1.5 were measured from $H_{160}$ COSMOS-DASH imaging (0.66 deg$^2$) and of 748 galaxies at z$<$1.5 from $I_{814}$ ACS-COSMOS imaging (1.7 deg$^2$) \citep{Koekemoer2007,Massey2009}. Photometric redshifts, stellar masses and rest-frame colors were taken from the UltraVISTA catalog \citep{Muzzin2013}.

In both surveys the sizes of galaxies were measured  by single-component S\`ersic profile fits to two-dimensional light distributions
using GALFIT \citep{Peng2010}, with a correction for redshift-dependent color gradients. Details  are described in \citet{VanderWel2014} and \citet{Mowla2018}. The two datasets have been combined carefully, verifying that there are no detected systematic differences between them. The combined size--mass distribution is the largest dataset with the largest ranges in stellar mass and redshift currently available, and is detailed in \citet{Mowla2018}. In this paper we only analyze data where we are mass-complete. The lower bounds of the stellar mass limits correspond to the mass-completeness limits down to which \citet{VanderWel2014} determined structural parameters for star-forming and quiescent galaxies with good fidelity.

\subsection{Galaxy size definition}

High redshift galaxies are typically modeled by single component S\`ersic profiles which describe the structure of a galaxy with the half-light radius $r_{50}$, the radius containing 50\% of light, and the S\`ersic index $n$, a measure of form of the light profile.  At a given stellar mass quiescent galaxies on average have a higher S\`ersic index and a smaller half-light radius than star-forming galaxies. As demonstrated in a companion paper (Miller et al.\ 2019), these two effects conspire such that the size difference between star forming and quiescent galaxies nearly disappears when using  $r_{\rm 80}$, the radius containing 80$\%$ of the stellar light. The size--mass relation is also tighter for this definition of radius, by approximately 0.06 dex (see Miller et al.\ 2019). Physically, this size definition is a better measure of the total baryonic extent, and it is in a regime where dark matter begins to dominate the mass. For a typical galaxy with $M_{\star}\sim 5 \times 10^{10}$ M$_{\odot}$ and dark matter halo mass $\sim 10^{12}$ M$_{\odot}$, the median fraction of dark matter contained within $r_{50}$ is 35\% to 55\% while that within $r_{80}$ is between 60\% to 80\%.

Following Miller et al.\ (2018), we calculate 
$r_{80}$ using the following relation:
\begin{equation}
    \frac{r_{80}}{r_{50}}= 0.0012 n^3\, -\, 0.0123 n^2\, +\, 0.5092n\, +\, 1.2646,  
\end{equation}
with $r_{50}$ the half-light radius and $n$ the S\`ersic index. 

\section{The Size--Mass Relation}

\subsection{Broken Power-Law Fit}
\begin{figure*}[htbp]
\centering
\includegraphics[width=0.98\textwidth]{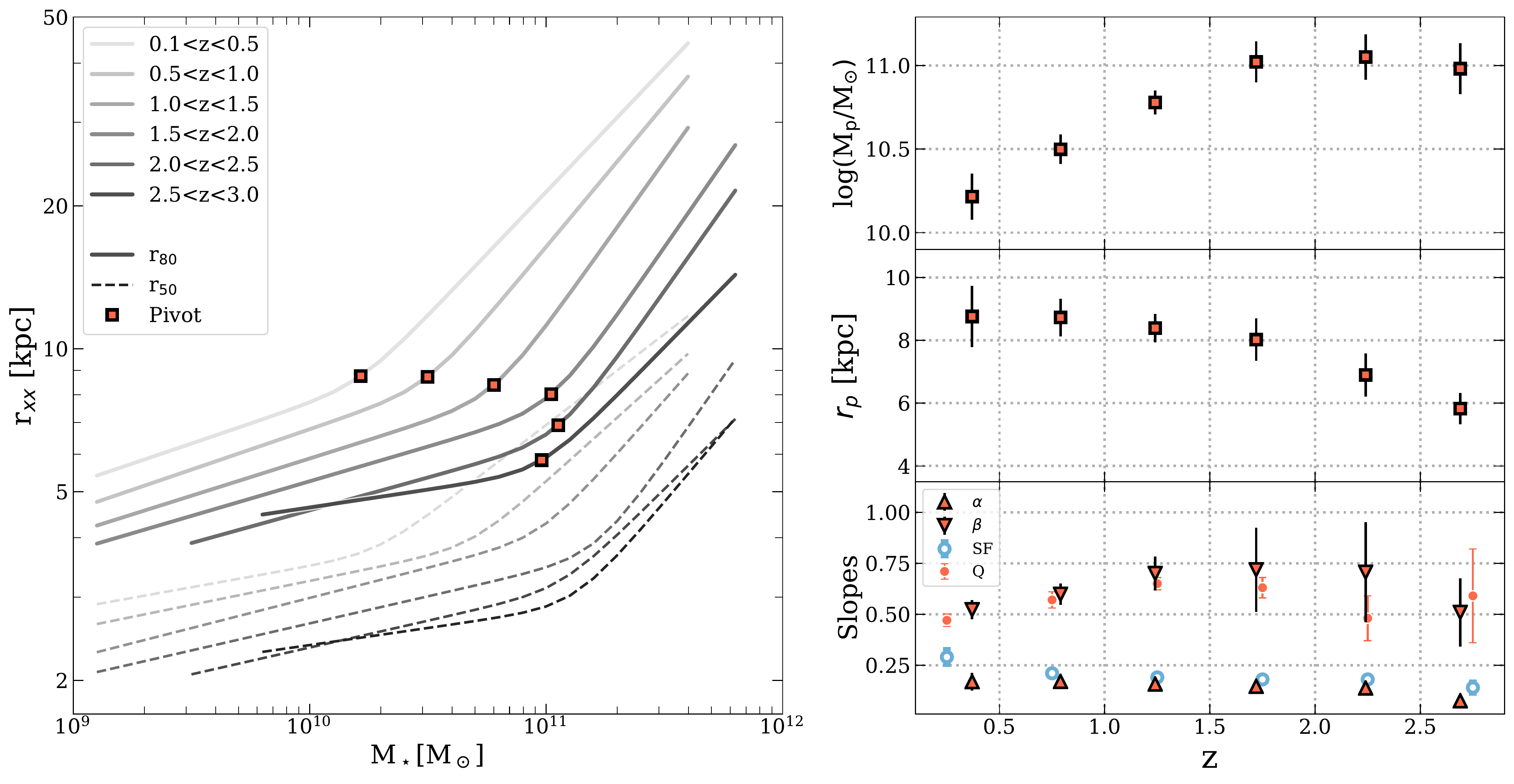}
\caption{Redshift evolution of smoothly broken power law fits (Eq. \ref{eq:power_law}) to the size--mass relation. \textit{Left:} Evolution of size--mass relation of all galaxies since z$\sim$3. The solid lines show the $r_{80}$ -- mass relation, while the broken lines show $r_{50}$ -- mass relation. The pivots of the broken power-law fits are indicated by red squares. \textit{Right:} Evolution of the parameters of broken power law fits to median $r_{80}$ -- mass distribution. The top panel shows the pivot stellar mass, the middle panel shows pivot radius and the bottom panel shows slopes $\alpha$ and $\beta$ of the power law. Overplotted are the slopes of the single power law fits to the star-forming and quiescent galaxies in \citet{Mowla2018}. } 
\label{fig:fit_evo}
\end{figure*}

The $r_{\rm 80}$ -- mass distributions of all galaxies with $\log(M_{\star}/{\rm M}_{\odot})>9$ in six bins of redshift are shown in Figure \ref{fig:data_median_fit}. The visible gaps in the distributions mark the points where the CANDELS sample \citep{VanderWel2014} transitions to the high mass ULTRAVISTA/COSMOS sample \citep{Mowla2018}. The median sizes of galaxies in mass bins are over-plotted on the size--mass distribution, which are color-coded by the fraction of galaxies which are quiescent; rest-frame U−V and V−J color space was used to separate galaxies into star-forming and quiescent. The error bars on median sizes are calculated as biweight scale divided by $\sqrt{N-1}$, where $N$ is the number of galaxies in each stellar mass bin. Visual inspection of the median size--mass relation shows that at the low-mass end the relation has a shallow slope, while at the high-mass end the relation steepens after a characteristic pivot point. Hence, we fit a smoothly broken power-law to the median size--mass relation of the form:

\begin{equation}
        r_{\rm 80} (M_\star) = r_p \left(\frac{M_\star}{M_p}\right)^{\alpha} \left[\frac{1}{2}\left\{ 1+\left(\frac{M_\star}{M_p}\right)^{\delta}\right\} \right]^{(\beta -\alpha)/\delta},
    \label{eq:power_law}
    \end{equation}

where $M_p$ is the pivot stellar mass at which the slopes change, $r_p$ is the radius at the pivot stellar mass, $\alpha$ is the slope at the low mass end, $\beta$ is the slope at the high mass end, and $\delta$ is the smoothing factor. We set the smoothing factor to $\delta=6$ to reduce degeneracy between $\delta$ and the slopes. We fit Eq. \ref{eq:power_law} to the median sizes at each redshift bin, using the `trust region reflective' algorithm as implemented in \texttt{curvefit} of \texttt{scipy}. The parameters of the best-fitting relations for all redshift bins are given in Table \ref{tab:fit}. 

\begin{table*}[htbp]
\centering
\caption{Best-fit parameters of smoothly broken power-law fit to the size--mass relation, given in Eq. \ref{eq:power_law}.}
\begin{tabular}{|c|cccc || cccc|}
\hline
\multicolumn{1}{|c}{\textbf{z}} & \multicolumn{4}{|c||}{\textbf{$r_{\rm 80}$}} & \multicolumn{4}{c|}{\textbf{$r_{50}$}} \\ \cline{2-9} 
 & $r_p$ [kpc] & log($M_{\rm p}/{\rm M}_{\odot}$) & $\alpha$ & $\beta$ & $r_p$ [kpc] & log($M_{\rm p}/{\rm M}_{\odot}$) & $\alpha$ & $\beta$\\ \hline
0.37 & 8.6$\pm$0.7 & 10.2$\pm$0.1 & 0.17 $\pm$ 0.03 & 0.50 $\pm$ 0.03 & 3.8$\pm$0.3 & 10.3$\pm$0.1 & 0.09 $\pm$ 0.03 & 0.37 $\pm$ 0.03   \\
0.79 & 8.7$\pm$0.5 & 10.5$\pm$0.1 & 0.17 $\pm$ 0.02 & 0.61 $\pm$ 0.04 & 4.0$\pm$0.4 & 10.7$\pm$0.2 & 0.10 $\pm$ 0.02 & 0.45 $\pm$ 0.09 \\
1.24 & 8.3$\pm$0.3 & 10.8$\pm$0.1 & 0.16 $\pm$ 0.01 & 0.69 $\pm$ 0.06 & 4.2$\pm$0.4 & 11.1$\pm$0.2 & 0.13 $\pm$ 0.01 & 0.53 $\pm$ 0.17 \\
1.72 & 7.6$\pm$0.7 & 10.9$\pm$0.1 & 0.15 $\pm$ 0.01 & 0.62 $\pm$ 0.19 & 3.7$\pm$0.8 & 11.1$\pm$0.5 & 0.11 $\pm$ 0.03 & 0.50 $\pm$ 0.29  \\
2.24 & 6.5$\pm$0.7 & 11.0$\pm$0.2 & 0.14 $\pm$ 0.02 & 0.53 $\pm$ 0.17 & 3.1$\pm$0.5 & 11.0$\pm$0.3 & 0.11 $\pm$ 0.02 & 0.42 $\pm$ 0.25  \\
2.69 & 5.3 $\pm$0.4 & 10.8$\pm$0.2 & 0.05 $\pm$ 0.03 & 0.34 $\pm$ 0.09 & 2.8 $\pm$0.4 & 10.9$\pm$0.3 & 0.06 $\pm$ 0.04 & 0.38 $\pm$ 0.20\\ \hline
\end{tabular}
\label{tab:fit}
\end{table*}

The fits are generally excellent, with reduced $\chi^2$ values ranging between 0.95 and 1.8. The most stable results are obtained in the redshift range $0.5<z<2.0$, as there are $>5000$ galaxies in each redshift bin with a large dynamic range in mass. At the lowest redshifts ($z<0.5$) the COSMOS field does not have sufficient volume to properly sample the full distribution; this may affect the characteristic pivot mass measurement. A similar problem arises at $z>2$, where the pivot stellar mass is high and we have a relatively low number of galaxies in the relevant mass range.  

\subsection{Redshift Evolution of the Size -- Stellar Mass Relation}

The parameters of the best-fit broken power law function are shown as a function of redshift
in the left panel of Fig. \ref{fig:fit_evo}. The best fitting functions are shown in the right panel, and are also overplotted in Fig.\ 1.  
We find that the pivot stellar mass decreased with cosmic time, going from $M_p = 8\times10^{10}\,{\rm M}_{\odot}$ at $z\sim1.72$ to $M_p = 1.4\times\,10^{10} {\rm M}_{\odot}$ at $z\sim0.25$. The evolution in pivot stellar mass appears to flatten off between z$\sim$1.5 and z$\sim$3; however, further study is required to investigate whether this is a physical phenomenon or due to small number statistics of high mass galaxies at z$>2$. In contrast to the pivot mass itself, the radius at the pivot mass increased with time, from $r_p=5.3$\,kpc at $z\sim2.75$ to $r_p=8.6$\,kpc at $z\sim0.25$.

The slope of the size--mass relation at $M_{\star}<M_p$ is approximately constant at $\alpha\approx 0.16$ while it is $\beta\approx 0.60$ at $M_{\star}>M_p$. We note here that the slope of the low mass end is similar to the slope of a single power law fit to the sample of star-forming galaxies, while that of the high mass galaxies is similar to a single power law slope of quiescent galaxies \citep[see][]{Mowla2018}. 

\begin{figure*}[htbp]
\centering
\includegraphics[width=\textwidth]{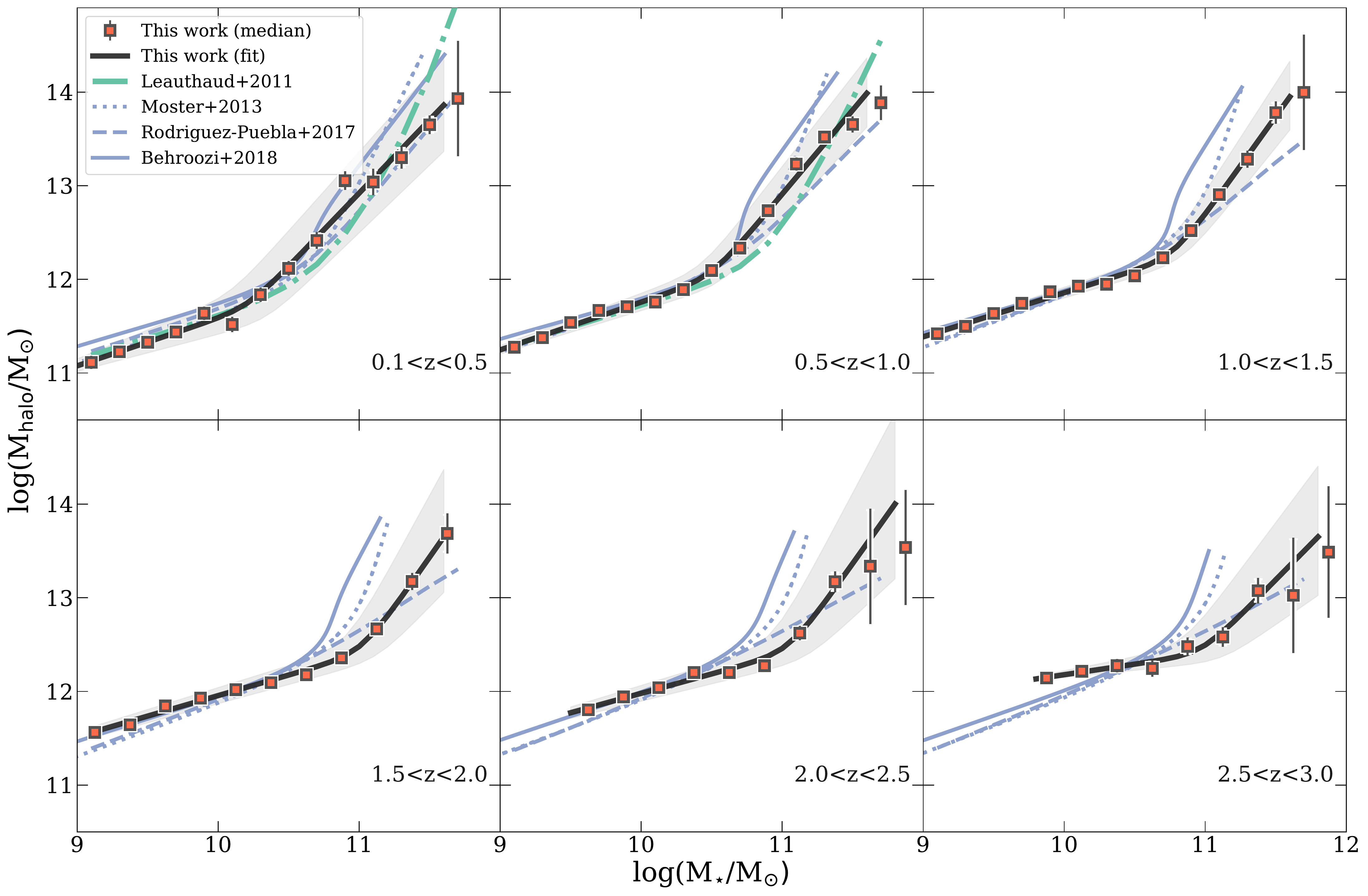}
\caption{Halo mass-stellar mass relation derived from the size--mass relation of galaxies at $0.1<z<3.0$. The halo masses are determined from virial radii, with the simple assumption that the virial radius scales as $R_{\rm vir} = \gamma^{-1}r_{80}$, with $\gamma = 0.047$. The red squares show the halo mass from median sizes and the black line shows the halo-mass-stellar mass relation from size--mass relation fit (grey band represents the error associated with the fit). Purple lines show stellar-to-halo mass relations derived using abundance matching techniques, and the green dashed line at $0.1<z<1.0$ is derived from galaxy-galaxy lensing and clustering \citep{Leauthaud2011}. } 
\label{fig:r80_mhalo}
\end{figure*}

\section{Halo-to-Stellar mass relation}
\label{sec:smhm}
\subsection{Calculating Halo Mass}
\label{sec:halo_calc}

The functional form of the size--mass relation is reminiscent of the form of the stellar mass -- halo mass relation: this relation also has different slopes in different mass regimes with an inflection point. In the SMHM relation the inflection point is where galaxy formation is maximally-efficient in the sense that the largest fraction of baryons is in stars \citep{Behroozi2010}. This superficial similarity motivates us to examine the hypothesis that the upturn in the stellar size -- stellar mass relation above the pivot stellar mass is simply a reflection of the downturn in the stellar mass -- halo mass relation above its pivot halo mass.  

\begin{figure*}[htbp]
\centering
\includegraphics[width=0.98\textwidth]{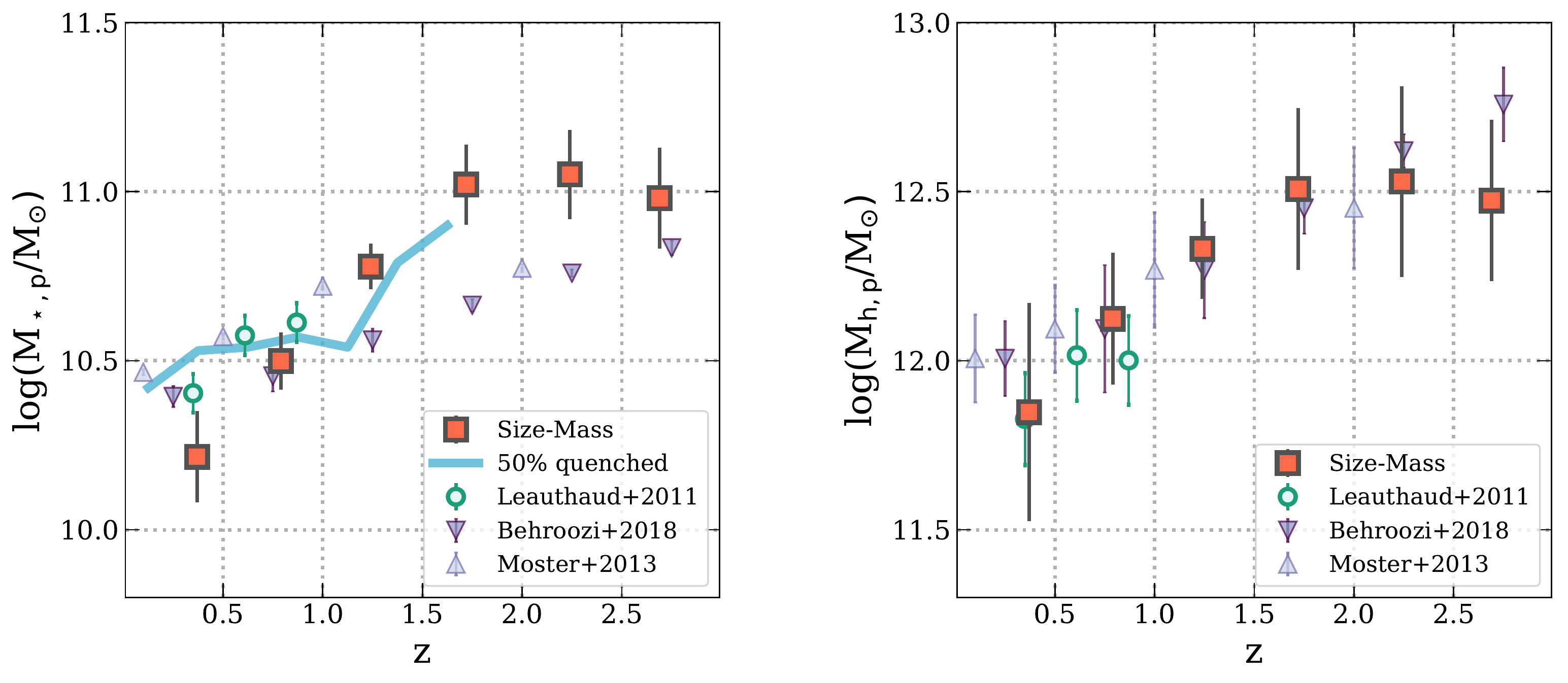}
\caption{Redshift evolution of pivot points. Left panel shows the evolution of pivot stellar mass of the broken power law fits to the size--mass relation (red circles), compared to similar fits to the stellar-to-halo mass relation from \citet{Behroozi2018}, \citet{Leauthaud2011} and \citet{Moster2013}. The blue-line marks the stellar mass at which 50$\%$ of the galaxies are quenched (in the UVISTA catalog). Right panel shows the calculated halo mass at pivot point of size--mass relation compared to pivot halo masses stellar-to-halo mass relation.} 
\label{fig:param_comp}
\end{figure*}

We test this by adopting a constant ratio between galaxy size and the virial radius of the halo: $R_{\rm vir}= r_{\rm 80}/\gamma$. We define the halo virial mass and virial radius within a spherical overdensity $\Delta_{\rm vir}$ times the critical density $\rho_{\rm crit}$: 
\begin{equation}
    M_{\rm halo} = \frac{4\pi}{3} \Delta_{\rm vir} \rho_{\rm crit} R_{\rm vir}^3,
    \label{eq:mhalo}
\end{equation}
where $\Delta_{\rm vir}$ is from \citet{Bryan1997StatisticalComparisons}.



This allows us to express median galaxy radii,
$r_{80}$, in terms of median halo masses, by choosing an appropriate value
for the proportionality constant $\gamma$. We fit for
$\gamma$ by minimizing the difference between the halo mass --
stellar mass relation that we derive in the lowest redshift
bin and the relation from \citet{Leauthaud2011} at $0.2<z<0.48$.
\citet{Leauthaud2011} measured halo masses from the COSMOS ACS data using a joint analysis of galaxy-galaxy weak lensing, galaxy spatial clustering, and galaxy number densities. We find $\gamma=0.047$ from this analysis. This value can be compared to previous studies that relate $r_{\rm 50}$ to $R_{\rm vir}$. These studies find $\gamma_{50}=0.015-0.03$ \citep{Kravtsov2012,Somerville2017,Huang2017,Jiang2018}. These values are consistent with our result for $r_{80}$, when the typical ratio between $r_{\rm 80}$ and $r_{\rm 50}$ is taken into account (a factor 2 to 3, depending on the Sersic index). 

The results are shown in Fig.\ 
\ref{fig:r80_mhalo}, and compared to halo mass -- stellar mass
relations from the literature. The derived halo-to-stellar mass function agrees very well with the lensing measurements
from \citet{Leauthaud2011} at all masses and both redshift ranges where lensing data are available, even though we fit only for a single offset. Beyond $z\sim 1$ we cannot compare directly to measurements, but as shown in \citet{Leauthaud2011} (and Fig.\ \ref{fig:r80_mhalo}) pivot halo mass measurements from lensing are consistent with those from halo occupation distribution (HOD) and subhalo abundance matching (SHAM) measurements. We therefore also include SMHM relations from SHAM and HOD measurements by \citet{Rodriguez-Puebla2017}, \citet{Moster2013}, \citet{Behroozi2018} and \citet{Legrand2018}. At all redshifts the SMHM that we derive from galaxy sizes agrees well with that derived using other methods, although slope of the high mass end and the pivot points start to diverge, as we will discuss later. 
This agreement with the much more sophisticated empirical modeling results is remarkable, especially given our simplistic assumption that, on average, $r_{\rm80}/R_{\rm vir} = 0.047$ across a wide range in stellar masses and cosmic epochs. In addition, the median halo mass at a given stellar mass is not necessarily equivalent to the median stellar mass at a given halo mass due to scatter in the relationship and the steepness of the mass function.

\subsection{The Pivot Mass}

A quantity that is of particular interest is the pivot mass, that is, the inflection mass where the slope of the SMHM changes.  To compare the pivot stellar masses between various SMHMs, we fit SMHMs from the literature with our smoothly broken power law relation (Eq.\ \ref{eq:power_law}) using the same methodology as the fits to the $r_{80}-M_*$ relation. The comparisons between the pivot stellar masses and pivot halo masses are shown in Fig.\ \ref{fig:param_comp}. The pivot points are in good agreement, although we see a stronger evolution of pivot stellar mass in the size--mass relation than in SMHM relations from abundance matching upto $z\sim3$. This has been noticed previously in \citet{Leauthaud2011} who finds a more significant evolution of pivot stellar mass in SMHM measured from lensing and clustering between redshift 0.2 to 1 than from SHAM measurements.  


\section{Discussion}

Using a new definition of size and the large galaxy sample from \citet{Mowla2018} 
we showed that the size--mass relation of all (quiescent plus star forming) galaxies is well fit with a broken power law. The stellar mass where the slope changes, the ``pivot mass", increases with redshift from $\log(M_{\rm p}/{\rm M}_\odot)\approx
10.2$ at $z=0.25$ to $\log(M_{\rm p}/{\rm M}_{\odot})\approx 11.0$ at $z=2.75$. We also showed that the form of this relation is remarkably similar to that of the stellar mass - halo mass (SMHM). The pivot stellar masses of the two relations are identical within the errors, and the slope and normalization are very similar when the simple scaling $r_{80}=0.047 R_{vir}$ is assumed for all masses and redshifts. As discussed in \S\,1, our results extend previous theoretical and observational studies \citep[e.g.,][]{Kravtsov2012,Somerville2017,Huang2017,Huang2018}. 

We note that our results do not rely on the use of $r_{80}$ instead of $r_{50}$; as shown in Fig. \ref{fig:fit_evo} we derive similar relations for $r_{50}$, although the change in slope is not as striking as it is for $r_{80}$. The main advantages of $r_{80}$ are that star forming galaxies and quiescent galaxies have similar sizes at fixed stellar mass (see Miller et al.\ 2019) and that it encompasses a larger fraction of the baryons.

It is interesting to speculate whether there is a straightforward
physical interpretation of the similarity of these relations.
From the stellar size--mass relation point of view, the pivot marks the stellar mass at which the galaxy population transitions from being dominated by star-forming galaxies  to being dominated by quiescent galaxies. 
This is shown explicitly by the blue line in Fig.\ 4, which indicates the mass where half the population is quiescent and half is star forming. This evolving mass matches the pivot mass within the errors, at least out to $z\sim 1.5$.
From the stellar-to-halo mass relation point of view, the pivot is where $M_{\star}/M_h$ reaches a maximum, i.e., it is the halo mass at which baryons have been most efficiently converted into stars. Taking these aspects together, the pivot may simply mark the mass above which both the stellar mass growth and the size growth transition from being star formation dominated to being (dry) merger dominated \citep[see also][]{Dekel2006}.

It is not immediately obvious why the pivot mass should evolve with redshift. However, following \citet{Leauthaud2011}, we note that the ratio of the pivot halo mass to the pivot stellar mass is roughly constant. That is, the evolution of the pivot mass and the size at the pivot mass conspire to keep the ratio $M_{\rm halo}/M_{\star}$ approximately constant at the pivot mass (at $\approx 0.025$).


Finally, we note that there are significant
caveats associated with inverting an average stellar-to-halo mass relation to an average halo-to-stellar mass relation, as is done in \S\,\ref{sec:smhm}. The existence of large, low surface brightness galaxies with very low velocity dispersions \citep{Danieli2019}, as well as the difference in clustering between star forming and quiescent galaxies of the same stellar mass \cite{Coil2017}, suggest that there is significant scatter in halo mass at fixed galaxy size.  As discussed in detail by \citet{Somerville2017} the scatter in the stellar-to-halo mass relation, combined with the exponential fall-off in the stellar mass function,  leads to an overestimate of the halo-to-stellar mass ratio at the high mass end. Indeed, the inverted \citet{Moster2013} and \citet{Behroozi2018} relations are steeper than our derived relation.  It is encouraging that our relation does agree with the direct estimates of \citet{Leauthaud2011}, who measure average halo mass at fixed stellar mass without the need of conversions or assumptions about the scatter. Turning this argument around, the fact that we see a clear break in our inferred halo mass -- stellar mass relation
may imply a small scatter in the stellar mass -- halo mass relation. We tested this by generating a mock catalog of galaxies using the SMHM from \cite{Behroozi2018} and introducing scatter in the stellar mass function. Beyond a scatter of 0.25 dex, the break in SMHM begins to disappear and the data can be reasonably well described by a single power law.  In this framework our inferred relation implies a scatter of no more than
0.2\,dex at the high mass end, in line with other constraints \citep[see][]{Moster2013,Behroozi2018}.



\acknowledgements

We thank Peter Behroozi, Andrey Kravtsov, Song Huang and Johannes Lange for valuable feedback. We also thank the anonymous referee for insightful comments which improved the manuscript. This work is supported by NASA HST program GO-14114. AvdW acknowledges funding through the H2020 ERC Consolidator Grant 683184.



\end{document}